# Magnetoactive elastomer based on superparamagnetic nanoparticles with Curie point close to room temperature

Yu.I. Dzhezherya[a,b,c], Wei Xu[d], S.V. Cherepov[c], Yu.B. Skirta[c], V.M. Kalita[b,c,e], A.V. Bodnaruk[e], N.A. Liedienov[a,f,*], A.V. Pashchenko[a,c,f,*], I.V. Fesych[g], G.G. Levchenko[a,f,*]

[a]*State Key Laboratory of Superhard Materials, International Center of Future Science, Jilin University, 130012 Changchun, China*
[b]*National Technical University of Ukraine "Igor Sikorsky Kyiv Polytechnic Institute", Kyiv 03056, Ukraine*
[c]*Institute of Magnetism, NAS of Ukraine and MES of Ukraine, Kyiv 03142, Ukraine*
[d]*State Key Laboratory of Inorganic Synthesis and Preparative Chemistry, College of Chemistry, Jilin University, Changchun, 130012, China*
[e]*Institute of Physics, NAS of Ukraine, Kyiv 03028, Ukraine*
[f]*Donetsk Institute for Physics and Engineering named after O.O. Galkin, NAS of Ukraine, 03028 Kyiv, Ukraine*
[g]*Taras Shevchenko National University of Kyiv, 01030 Kyiv, Ukraine*

*Corresponding authors:
E-mail address:  nikita.ledenev.ssp@gmail.com (N.A. Liedienov),
alpash@ukr.net (A.V. Pashchenko),
g-levch@ukr.net (G.G. Levchenko)

**Abstract**

A magnetoactive elastomer (MAE) consisting of single-domain $La_{0.8}Ag_{0.2}Mn_{1.2}O_3$ nanoparticles with a Curie temperature close to room temperature ($T_C$ = 308 K) in a silicone matrix has been prepared and comprehensively studied. It has been found that at room temperature and above, MAE particles are magnetized superparamagnetically with a low coercivity below 10 Oe, and the influence of magnetic anisotropy on the appearance of a torque is justified. A coupling between magnetization and magnetoelasticity has been also established. The mechanisms of the appearance of magnetoelasticity, including the effect of MAE rearrangement and MAE compression by magnetized particles, have been revealed. It has been found that the magnetoelastic properties of MAE have critical features near $T_C$. The magnetoelastic properties of MAE disappear at $T > T_C$ and are restored at $T < T_C$. This makes it possible to use MAE at room temperature as a smart material for devices with self-regulating magnetoelastic properties.

# 1. Introduction

The study of composites with magnetic micro- or nanoparticles located in a polymer matrix is of great interest due to the possibility of remote (contactless) control of their properties by magnetic field and their application in soft robotics, biomedical, civil engineering, [1-6]. If the particles are in an elastomer matrix, they can be displaced relative to each other under the magnetic interparticle interaction forces [7-10]. It leads to deformation of the matrix and the appearance of composite restructuring under magnetic field, and, as a consequence, the particle chains or a columnar structure can be formed. Composites consisting of an elastomer matrix and an ferromagnetic particles are called magnetoactive elastomers (MAE) or magneto-rheological elastomers (MRE) [11].

MAE restructuring leads to several interesting phenomena, which are developed atypically in comparison with solids, i.e. an anomalous MRE [12-19], magneto-dielectric effect [20-22], magnetic shape memory [23-27]. Taking into account the significant response or flexibility of such materials on external influences, MAE is referred to as smart materials [28]. In the literature, MAE with microparticles is most often studied [1,11,15,29]. Nanoparticles are much smaller, so that it is not entirely clear how the restructuring of MAE with nanoparticles occurs.

Earlier MAEs with magnetic nanoparticles were studied, for example, in [30-38]. In comparison with MAE consisting of magnetic microparticles, it turned out that the MRE for MAE with nanoparticles is weaker [29,39]. For example, the hematite nanoparticles used in MAE have a lower magnetization compared to the magnetization of microparticles of the transition metal group as well as due to the size effect [40,41]. Therefore, it becomes more difficult for a weakly magnetized filler to achieve interparticle magnetic forces that can exceed the elastic forces of interparticle interactions. However, despite the low magnetization, it can be expected that due to their smaller size, the nanoparticles during their magnetization can be more mobile in the polymer matrix, which should affect the MRE.

In addition, nanoparticles compared to microparticles have a completely different magnetization mechanism. Ferromagnetic microparticles are in a multidomain state. The nanoparticles with sizes less the critical one are single-domain [42] and magnetized in a magnetic field as the coherent rotation of the magnetic moments of particle ions [43]. At high temperatures, nanoparticles are magnetized superparamagnetically [44,45]. The behavior of their magnetization is determined by the thermoactivation mechanism, has Langevin form [46], and does not follow, for example, Rayleigh's law [16]. The magnetic susceptibility of an ensemble of superparamagnetic particles can be large, despite the small value of their saturation magnetization. All of the above listed MAE features are associated with their magnetoelasticity.

It is assumed that the magnetoelastic effect of the magnetic filler should disappear above the Curie temperature after the transition of the filler particles to the paramagnetic state. The possibility of observing this effect in MAE and its application depends on the Curie temperature of the magnetic filler. If the Curie temperature is significantly higher or lower than room temperature, the practical use of this phenomenon in MAE is not possible. However, there are compounds, where the Curie temperature is relatively easily changed by controlling their stoichiometry or doping. These are, for example, manganites [47-52]. It is possible to synthesize manganites with a Curie temperature close to room temperature or slightly above. At the Curie point, the behavior of a magnet is anomalous, which is also interesting both from the point of view of their fundamental study and practical application. Compared to metal particles, manganite particles are more convenient and promising, for example, for medical applications [53-56].

The possibility of switching off the magnetic-elastic interactions by changing temperature are relevant and this effect can be additional controling parameter of these smart materials. This temperature effect can be similar to the solidification or glass transition of the MAE matrix upon cooling [9,10], when the MAE matrix becomes more rigid, the position of particles in the matrix becomes blocked, and restructuring of the MAE stops. However, near $T_C$, the effect of weakening magnetoelastic coupling is associated not with a change in the elastic properties of the matrix, but with a change in the magnetic properties of the particles. When the temperature drops below the Curie point, the magnetoelastic effect turns on again. Thus, beside the critical temperatures of the MAE matrix and in the case of using nanoparticles with a Curie temperature below the flow point of the matrix polymer (the transition temperature between the highly elastic and viscous-flow state), an additional opportunity arises for manipulating the magnetoelastic properties of the MAE. This process will occur as a self-developing or self-controlled process.

For modern soft robotics, the study of bending deformations of MAE is relevant. A special interest in the MAE bending induced by the magnetic field is also associated with designing and creating 3D magnetoactive metamaterials for acoustics [57]. The modeling of the MAE bending induced by the magnetic field is a nontrivial mathematical problem [58, 59]. Moreover, the bending and restructuring of the MAE with nanoparticles have not been enough studied.

In this work, we have investigated the magnetic and magnetoelastic properties of MAE, which consists of superparamagnetic nanoparticles with a Curie point of about room temperature, placed in a highly elastic silicone matrix.

The main aspects of this paper are the study of: (i) the functional properties of the MAE; (ii) the bending deformations of the MAE beam induced by the magnetic field; (iii) the interrelation between the bending and magnetization of the MAE; (iv) the features of the magnetoelastic coupling of MAE with superparamagnetic nanoparticles during heating to the Curie point.

## 2. Experimental

The manufacturing of the MAE traditionally consists of two stages: (i) mixing and (ii) hardening. To fabricate MAE matrix, we used the magnetic $La_{0.8}Ag_{0.2}Mn_{1.2}O_3$ nanopowder with Curie temperature $T_C = 308K$ and two-component of soft silicone compound Silikon SKR-780 (Slovakia) with a density of 1.1-1.15 g/cm$^3$ at 25 °C, a hardness of 5 units (Shore A), an elongation to break of 600%. Average viscosity at 25 °C is 15000 CPS. Conditional tensile strength is 2.4 MPa. Tear resistance is 14 kN/m. The MAE density is $\rho = 1.60$ g/cm$^3$, and the concentration of manganite particles is $v = 33$ mass %. From measurements, Young's modulus of the studied MAE sample is equal to 0.57 MPa. The La0.8Ag0.2Mn1.2O3 nanopowder was chosen because its Curie temperature is close to room temperature and can be easily changed by doping or changing stoichiometry.

The magnetic $La_{0.8}Ag_{0.2}Mn_{1.2}O_3$ nanopowder was synthesized using the pyrolytic decomposition of nitrates method [60]. The formation of a rhombohedral $R\overline{3}c$ perovskite $La_{0.8}Ag_{0.2}Mn_{1.2}O_3$ crystal structure was confirmed using Shimadzu LabX XRD-6000 X-ray diffractometer in Cu$_{K\alpha}$-radiation, $\lambda = 0.15406$ Å. The X-ray density of the $La_{0.8}Ag_{0.2}Mn_{1.2}O_3$ nanopowder is $\rho = 6.213$ g/cm$^3$.

The $La_{0.8}Ag_{0.2}Mn_{1.2}O_3$ manganite powder consists of spherical-like nanoparticles with a size of 65 nm (see Fig. 1(a)), which was determined using scanning electron microscopy (SEM) on an FEI MAGELLAN 400 Scanning Electron Microscope and JEOL JEM-2200FS transmission electron microscopy (TEM). The lattice interplanar distance of the $La_{0.8}Ag_{0.2}Mn_{1.2}O_3$ nanoparticles (see Fig. 1(b)) was also obtained using high-resolution TEM (HRTEM) with accelerating voltage of 200 kV. For that, the Fast Fourier Transform (FFT) using Gatan Microscopy Suite software was employed (see insert in Fig.1(b)). Fig. 1(c) demonstrates the corresponding lattice plane intensity profile with the interplanar distance of 0.377 nm (012) obtained from FFT, which is in a good agreement with the XRD data.

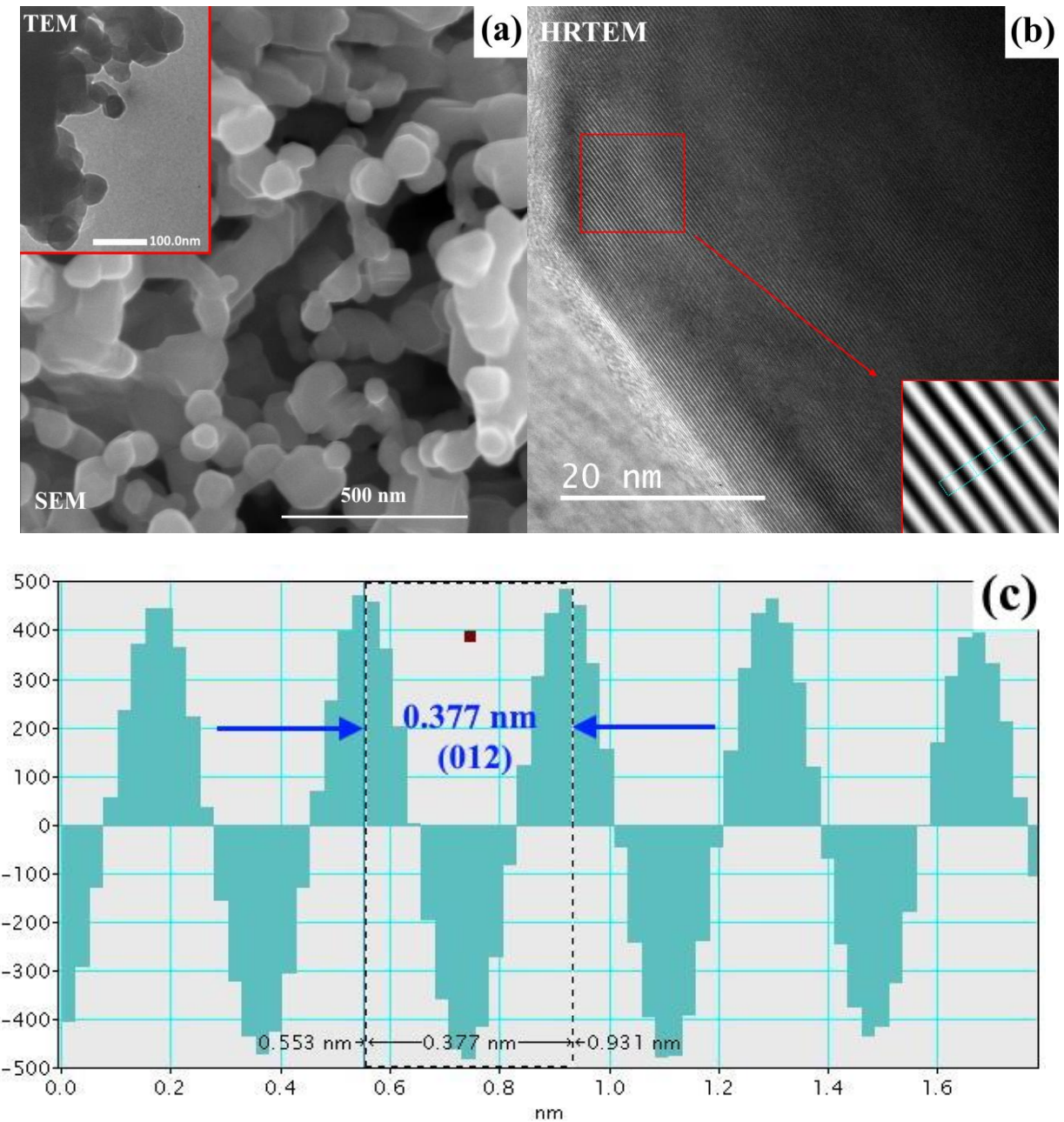

**Fig. 1**. Structural characterization of the $La_{0.8}Ag_{0.2}Mn_{1.2}$ nanopowder: (a) SEM (the insert shows TEM), (b) HRTEM (the insert shows FFT), and (c) lattice plane intensity profile corresponding to (012) plane.

The magnetic measurements were performed using Quantum Design SQUID MPMS 3 and vibration magnetometer LDJ 9500 in a temperature range from $T = 77$ to 400 K and in magnetic field up to $H = 7$ T. The value of magnetization was measured per unit mass of the $La_{0.8}Ag_{0.2}Mn_{1.2}O_3$ powder or per unit mass of the MAE.

The magnetoelastic properties of the MAE and the features of the magnetoelastic coupling near the Curie point were studied in course of the bending of the MAE beam induced by the magnetic field. Dimensions of the beam are $a = 5$, $b = 20$, $c = 1$ mm.

Scheme for measuring bending of MAE beam is shown in Fig. 2. The sample was placed between the poles of an electromagnet connected to a DC source. The diameter of the electromagnet poles is 165 mm, and the distance between them is 82 mm, which provides a sufficiently high homogeneous of the magnetic field, where the sample was located. The MAE beam 1 was placed equidistantly between the poles of the electromagnet in the sample holder 2 and perpendicularly to the magnetic field. One edge of the sample was fixed in clamp 3, and other was free. A displacement of the free edge of MAE beam along the field characterizes its bending and is determined by $\Delta L$ (see Fig. 2). This scheme makes it possible to measure the magnitude of the bending $\Delta L$ under the magnetic field and heating the sample.

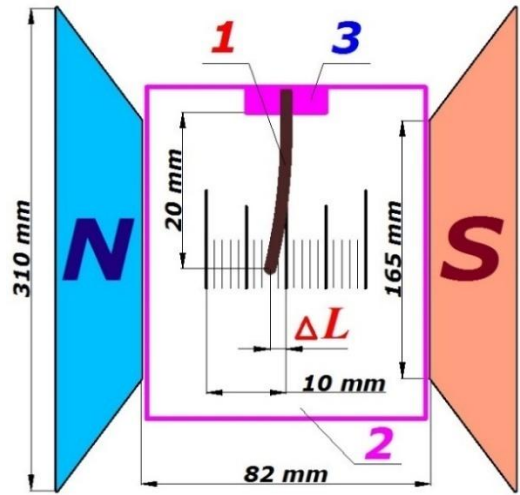

**Fig. 2**. The scheme for measuring a MAE beam bending (top view). The sample is between the poles of an electromagnet *NS*. The position of one edge of the sample is fixed and the other is shifted by $\Delta L$.

## 3. Results and discussion

### 3.1. Comparative analysis of magnetic properties between $La_{0.6}Ag_{0.2}Mn_{1.2}O_3$ nanoparticles and MAE

As can be seen from the hysteresis loops for both MAE and nanopowder (see Fig. 3), the particles of nanopowder are in PM state at high temperatures, when magnetization is proportional to the magnetic field ($m \sim H$), and FM state with a saturated magnetization in a strong magnetic field.

At $T = 300$ K, the magnetization under $H = 10$ kOe is $m = 35.2$ emu/g and 10.8 emu/g for the powder and MAE, respectively. From the ratio of these values, the percentage of FM particles in the MAE sample is 30.5%, which is in good agreement with the mass fraction of nanoparticles used for preparation of the MAE. The decrease in the magnetization of nanoparticles in the MAE is insignificant.

From the hysteresis loops for the powder (see bottom inset in Fig. 3), the coercivity is $H_C = 140$ Oe at $T = 77$ K and 55 Oe at $T = 300$ K. For MAE at $T = 290$ K, the coercivity is $H_C = 10$ Oe, which is much less than for the powder. In MAE, the particles are located less densely (at a greater distance from each other) than in the powder, and the interparticle magnetic dipole interaction between them is weakened, which leads to a decrease in the coercivity. With such a low coercivity, it can be argued that the ensemble of nanoparticles in the MAE are magnetized by equilibrium and superparamagnetic ways almost in the entire range of magnetic field (except for a very narrow region near $H = 0$).

As it can be seen from the bottom inset in Fig. 3, the magnetization of nanopowder is not saturated even in a strong field ($H = 70$ kOe) at room temperature, which is below the Curie point (see Fig. 4). At $T = 77$ K, the saturation of the magnetization is well achieved even in low fields. However, the saturation magnetization depends on temperature (see Fig. 3). Therefore, the superparamagnetic magnetization of an ensemble of particles is not a strict Langevin one [61]. In Langevin particles, the magnetic moment modulus does not depend on field and temperature. In our case, in the vicinity of $T_C$, the magnetic moment of the particle is not constant and depends on the temperature and field.

The top inset in Fig. 3 shows the Arrot curves [62] of $m^2(H/m)$, which proves a phase transition of the particles in MAE. It can be seen that the particles behave in a magnetic field as a ferromagnetic at $T = 306$ K and below, and as paramagnetic at $T = 310$ K and above [60].

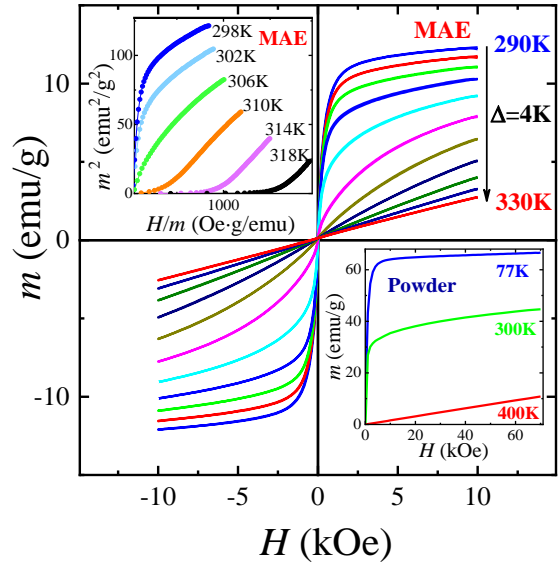

**Fig. 3**. Field dependences of magnetization for MAE with $La_{0.6}Ag_{0.2}Mn_{1.2}O_3$ nanoparticles near the Curie point. The top inset shows the Arrot's curves for MAE, and the bottom inset shows the upper parts of the hysteresis loops for $La_{0.6}Ag_{0.2}Mn_{1.2}O_3$ powder.

FC temperature dependences $m_{FC}(T)$ for determination of the Curie temperature in the powder and MAE, obtained in a constant magnetic field, were measured (see Fig. 4). For the powder and MAE, the measurements were carried out in $H = 50$ Oe and 2 kOe, respectively. Both dependences demonstrate a rapid change in the magnetization in the vicinity of the Curie point at $T \to T_C$. The field $H = 50$ Oe used for obtaining the $m_{FC}(T)$ curves is less than the value of the coercivity for the powder at room temperature. Therefore, it should make sure that the course of the $m_{FC}(T, H = 50$ Oe$)$ curve is not affected by blocking the directions of the magnetic moments of particles due to the field of their magnetic anisotropy [60,63]. Since the field $H =$ 2 kOe is much greater than the value of the coercivity of the particles, the effect of blocking the directions of the magnetic moments of particles by their magnetic anisotropy field will not be observed. As seen from Fig. 4, the temperature dependence of the $m_{FC}(T)$ dependence for the MAE in the vicinity of $T_C$ becomes flatter than for the powder. However, the field $H = 2$ kOe is not so large in comparison with the exchange field acted inside the particle to influence the critical behavior of the magnetization near the Curie point [64].

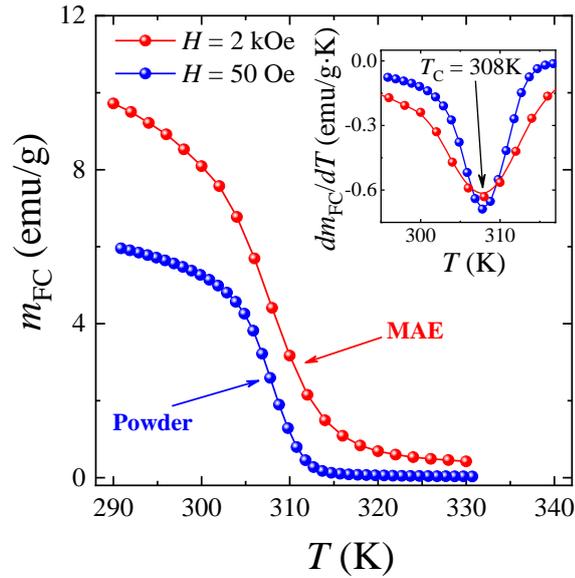

**Fig. 4**. The temperature dependences of FC magnetization $m_{FC}(T)$ for the powder and MAE measured under magnetic field $H = 50$ and 2 kOe, respectively. The inset shows the temperature dependences of the $dm_{FC}(T)/dT$ for determination of Curie temperature $T_C$ for the MAE and powder.

The Curie temperature has been obtained from the temperature dependences of $dm_{FC}(T)/dT$ for the MAE and powder (see inset in Fig. 4). Both $dm_{FC}(T)/dT$ curves, despite the fact that they were obtained in different fields, have a minimum at the same point $T_C = 308$ K. This value of the Curie point also coincides with the data of magnetocaloric effect for the same powder [60]. The Curie temperature $T_C = 308$ K for the MAE coincides with the $T_C$ for the powder. This manifests that magnetic anisotropy field of the particles is small and does not affect the $T_C$.

Thus, the chemical contact and the matrix do not significantly affect the magnetic properties of particles in the MAE. The coercivity of particles in the MAE decreases due to weakening of the mutual influence of particles on each other during their magnetization. Near the Curie point, the nanoparticles in MAE are magnetized superparamagnetically with a small value of the coercivity that is associated with the finite time of their Neel relaxation [44]. As it has been found out recently [60], the contribution to the magnetocaloric effect from SPM in the $La_{0.6}Ag_{0.2}Mn_{1.2}O_3$ nanoparticles is proportional to the square of the magnetic field, and its critical point is slightly less than $T_C$ and is equal to 306 K. As it has been turned out, the same feature in the $T_C$ region for the temperature dependence of the MAE bending with $La_{0.6}Ag_{0.2}Mn_{1.2}O_3$ nanoparticles induced by the magnetic field is obtained, the results of which are presented below.

### 3.2. Field dependences of the MAE bending

Fig. 5 shows the MAE bending depending on the magnitude and direction of a magnetic field. The field sweep was carried out gradually with an exposure time at each point for about 200 s. The cycle of changing magnetic field is: (i) an increase up to $H$ = 1650 Oe; (ii) a decrease to $H$ = 0 Oe; (iii) a change of direction of $H$ and an increase up to $H$ = -1650 Oe; (iv) a decrease to $H$ = 0 Oe (see arrows in Fig. 5). The lowest curve is the first input of the magnetic field. There is a residual bending when the magnetic field is turned off. The magnitude of the bending does not change its direction with the subsequent introduction of the magnetic field. A change in the direction of the bending with a change in the direction of the magnetic field would mean the appearance of an effect of similar to piezomagnetism, which is excluded in the case of equilibrium magnetization of the MAE. However, a change in the sign of bending can occur, for example, during magnetization of the MAE when there is a large remnant magnetization directed along the axis of the beam.

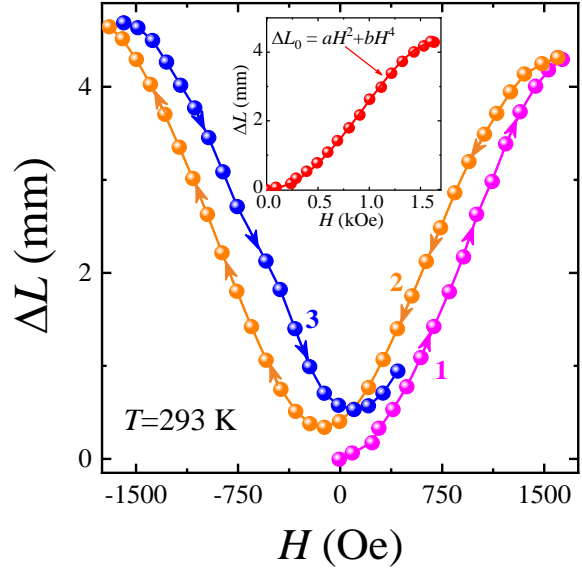

**Fig. 5**. Field dependence of the MAE bending in the cycle: $H = 0$ Oe → 1650 Oe (1), 1650 Oe → 0 Oe → -1650 Oe (2), -1650 Oe → 420 Oe (3). The direction of the field sweep is shown by arrows. The inset shows the initial bending and its approximation by a polynomial.

The presented hysteresis for the MAE bending in Fig. 5 is qualitatively very similar to the hysteresis for the elongation of a cylindrical MAE with microparticles of carbonyl iron during magnetization (magnetostriction) [65]. The occurrence of permanent deformation can be associated with the plasticity of the matrix. It should be noted that the MAE restructuring in a magnetic field is also a nonequilibrium process, and it can be another reason of plasticity upon deformation of the MAE by an external magnetic field. In the general case, plasticity and remnant deformation are a consequence of the non-equilibrium characteristic of MAE under external influences associated with the nonequilibrium process of deformation of the matrix [65] and the nonequilibrium process of displacement in the matrix of magnetized and interacting particles [10,16].

The curve of the first input of the magnetic field is closest to equilibrium state. The inset in Fig. 5 shows the field dependence for the bending upon the first input of the magnetic field, and the solid curve indicates the approximation of $\Delta L_0(H)$ in the field interval of $H$ = 0–1.65 kOe by the $\Delta L(H) = aH^2+bH^4$ polynomial, where $a$ = 3.1752 mm/kOe$^2$, $b$ = -0.587 mm/kOe$^4$. This approximation excludes the "piezomagnetic" effect in the studied MAE sample, and the inversion of the $H \rightarrow -H$ field occurs without changing the direction (sign) of bending $\Delta L$.

### 3.3. Temperature dependence of bending

Fig. 6 shows the temperature dependence of the beam bending $\Delta L(T)$ while its heating in a magnetic field $H$ = 2kOe, when the bending and magnetization are close to saturation. The heating rate is 0.05 K/s. With increase in a temperature, the bending decreases and finally disappears at $T > T_C$. This is due to the fact that the bending is caused by magnetoelastic forces and it disappears as the magnetization decreases. The temperature variation of the bending value has a critical behavior with a critical temperature $T_{cr}$ = 306 K (see the inset in Fig. 6). The derivative of $d\Delta L/dT$ has a minimum at $T = T_{cr}$.

There is no residual bending after passing the Curie point. This result confirms the above assumption that plasticity is associated with MAE restructuring. Above the Curie point, the particles become PM and weakly magnetized. Therefore, there is no restructuring due to their interaction with each other, and, as a consequence, the residual bending at a temperature above the Curie point tends to zero, $\Delta L(T > T_C) \rightarrow 0$.

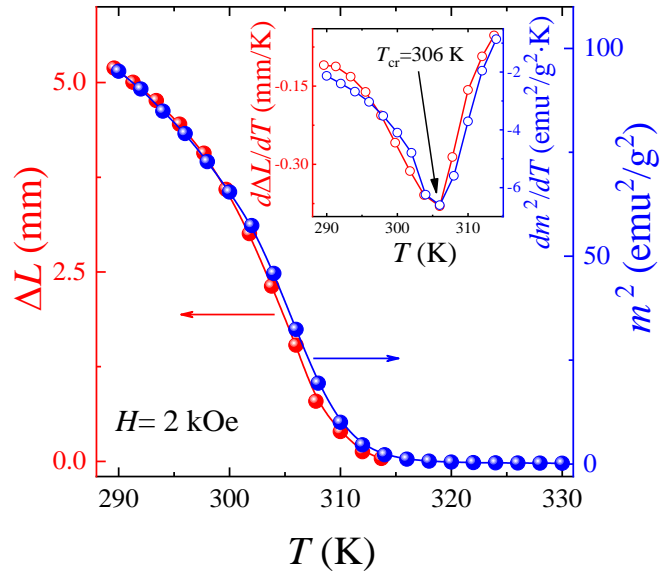

**Fig. 6**. Temperature dependence of the bending $\Delta L(T)$ and the square of the magnetization $m^2(T)$ of the MAE in $H = 2$ kOe. The inset shows the derivatives of $d\Delta L/dT$ and $dm^2(T)/dT$, which have a minimum at the point of $T_{cr} = 306$ K.

For comparison with $\Delta L(T)$, the temperature dependence of the square of the magnetization $m^2(T)$ in a magnetic field $H = 2$ kOe is shown in Fig. 6. It can be argued that the magnitude of the bending is directly proportional to the square of the magnetization of the MAE, $\Delta L(T, H=2\text{kOe}) / \Delta L(T = 290 \text{ K}, H=2\text{kOe}) = m^2(T, H=2\text{kOe}) / m^2(T = 290 \text{ K}, H=2\text{kOe})$. It should be noted that magnetostriction in ferromagnets may be a more complex dependence on magnetization [66, 67].

The inset in Fig. 6 shows the temperature dependence of the derivative of the square of the magnetization $dm^2(T)/dT$. As it can be seen, the temperature dependence of the derivative $dm^2(T)/dT$ has a minimum at the point of $T_{cr} = 306$ K.

From Fig. 6 can be seen that temperature dependences of $d\Delta L/dT$ and $dm^2(T)/dT$ have the same temperature of the minimum points and behave similarly near the minimum point $T_{cr} = 306$ K. This position of the minimum point is slightly less than the Curie point of the manganite particles $T_C = 308$ K (see Fig. 4). The shift of the critical point for bending is due to the fact that the value of $\Delta L$ is directly proportional to the square of the magnetization. The similar shift in the critical temperature is also observed for the superparamagnetic contribution to the magnetocaloric effect of the $La_{0.6}Ag_{0.2}Mn_{1.2}O_3$ manganite nanoparticles [60].

### 3.4. Magnetoelastic mechanisms of MAE bending in a magnetic field

The thermodynamic approach based on the elastic and magnetoelastic energies of the MAE in the approximation of a small bending $\varepsilon = \Delta L/l \ll 1$, where $l$ is the beam length, was used to describe the MAE bending in a magnetic field. The elastic energy of the beam is positive and is equal to $E_{el} = g\varepsilon^2/2$, where $g$ is the stiffness coefficient of the beam upon bending. The beam acquires this energy due to the work of $A_H$ performed by the moment of forces created by the external magnetic field. The $A_H$ for small deformations can be written as $A_H = kmH\varepsilon$, where $k$ is the coefficient. In this expression, it is taken into account that the moment of forces is proportional to the magnetization and the field strength, and at $\varepsilon \ll 1$, the value of the $A_H$ is proportional to the displacement. The $A_H$ is performed by the energy of the magnetic field $E_{mag} = -A_H$. In equilibrium state, the total energy $E_{total} = E_{el} + E_{mag}$ should be minimal as $dE_{total}/d\varepsilon = 0$, where we obtain the equation $g\varepsilon - kmH = 0$. It satisfies the condition of mechanical equilibrium: the magnetic and mechanical moments of forces applied to the beam compensate each other. From this equation we find that $\varepsilon = kmH/g$ and deformation occurs only in a magnetic field when the MAE is magnetized. In low fields, the magnetization of superparamagnetic particles is directly proportional to the magnetic field $m = \chi_H H$, where $\chi_H$ –effective (sample-averaged) magnetic susceptibility of superparamagnetic particles in the MAE. Thus, the deformation in low fields is quadratic to the magnitude of the magnetic field $\varepsilon = 2k \cdot \chi_H H^2/g$ (see Fig. 6) [68]. Phenomenological theory shows that during bending of the MAE, the magnitude of deformation is an even field effect.

The bending of a weakly coercive MAE with nanoparticles may be due to magnetic inhomogeneity. This can be inhomogeneity of the magnetic field and/or inhomogeneity of the magnetic properties of the MAE caused by restructuring. The bending can be caused by the magnetic anisotropy of the MAE sample, which magnetizes in different ways depending on magnetic field direction.

Let us consider the elementary MAE volume $dV$ with a magnetization **m** which is proportional to the field $\mathbf{m} = \chi_H \mathbf{H}$ in a small magnetic field **H**, which is perpendicular to the beam. The magnetic energy $dW$ of this MAE section will be equal to $dW = -\frac{1}{2}\mu_0 \chi_H H^2 dV$. The magnitude of the force acting on this section d$V$ is quadratic in the field and is equal to $d\mathbf{F} = \frac{1}{2}\mu_0 \nabla(\chi_H H^2)dV$. It should be noted that the inhomogeneous field of $\mathbf{H}(\mathbf{r}) \neq$ const displaces the MAE, and the inhomogeneous magnetization, when the local $\chi(\mathbf{r}) \neq$ const, leads to internal compressions / stretches of the MAE.

According to the generalized Gauss-Ostrogradsky theorem, this volume integral can be written as a surface integral $\Delta \mathbf{F} = \frac{1}{2}\mu_0 \oint (\chi_H H^2) d\mathbf{S}$, where $\Delta \mathbf{F}$ acts on the selected part of the MAE, the volume of which is $\Delta V$. For a flat sample surface in an external magnetic field perpendicular to this surface, it is possible to write $\Delta \mathbf{F} = \frac{1}{2}\mu_0 \Delta S \chi_H (H^2(\delta/2) - H^2(-\delta/2))\mathbf{n}$, where $\Delta S \delta = \Delta V$, $\mathbf{n}$ – unit vector (see Fig. 7 (a)) and $H(\pm\delta/2)$ are the values of the magnetizing field at the edges of the surface with coordinates $\pm\delta/2$. If the magnetic field is not uniform $H(-\delta/2) \neq H(\delta/2)$, the force acting on the MAE from the external magnetic field will not be zero $\Delta \mathbf{F} \neq 0$. Fig. 7 (b) shows the force vector $\Delta \mathbf{F}$ acting on the MAE section with volume $\Delta V$ for $H(\delta/2) > H(-\delta/2)$,

The influence of a weakly inhomogeneous magnetic field can be increased due to additional inhomogeneity in the arrangement of particles if they have the ability to shift into the MAE under a magnetic field, i.e. due to the restructuring of the MAE [7-10]. The magnetized nanoparticles will move in the matrix toward the sample surface along the direction of the greatest increase in the magnetic field. Therefore, near the surface of the sample, where the magnetic field strength is greatest, the concentration of particles and the magnitude of the magnetic susceptibility are both increased. An increase in the concentration of magnetized particles will create an additional stress along this surface of the sample, compressing the MAE. Compression of the MAE in the near-surface layer will promote bending of the sample. Thus, the restructuring can lead to an increase in the magnitude of the bending of the MAE beam. The bending of the beam can be large even in a small and weakly inhomogeneous magnetic field.

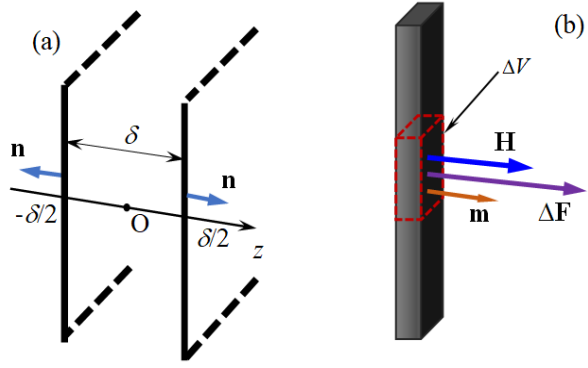

**Fig. 7**. Cross-section of the MAE sample with a thickness of $\delta$; $z$ is the coordinate axis, the origin of which is located in the middle of the sample; $\pm\delta/2$ are coordinates of the edges of flat surfaces; **n** are unit vectors to these surfaces (a). Magnetization **m**, magnetic field **H**, and force $\Delta\mathbf{F}$ acting on a section of the beam with volume $\Delta V$ (b).

The restructuring of particles in MAE occurs as a nonequilibrium process and has hysteresis in the change in position even for a pair of magnetized spherical particles [69,70]. In MAE with nanoparticles, the bending hysteresis (see Fig. 5) induced by the magnetic field can also be a consequence of a nonequilibrium change in the position of nanoparticles in the matrix due to magnetic dipole-dipole interactions between them.

Field dependences of the magnetizations $m_\parallel(H_\parallel)$ and $m_\perp(H_\perp)$ for the MAE beam with fixed edges are presented in Fig. 8 when the magnetic field is directed along the axis of the beam ($H_\parallel$) and perpendicular to the axis of the largest plane of the beam ($H_\perp$) as in Fig. 7(b). These dependences demonstrate the effect of the sample shape anisotropy on magnetization. The influence of the shape leads to the fact that the field in the middle of the MAE depends on the average magnetization of the sample and is less than the external magnetic field. This is an additional source of nonlinear magnetization of the MAE sample with superparamagnetic nanoparticles [71].

The axis along the beam is the axis of easy magnetization $\chi_\parallel > \chi_\perp$ that is well shown in the inset of Fig. 8 as field dependences for the derivatives of $dm_\perp/dH_\perp$ and $dm_\text{II}/dH_\text{II}$. At $H \to 0$, the value of the $\chi_\parallel$ is almost 50% higher than the $\chi_\perp$. The value of $dm_\perp/dH_\perp$ at $H_\perp \to 0$ coincides with the average effective $\chi_H$ susceptibility of the MAE sample.

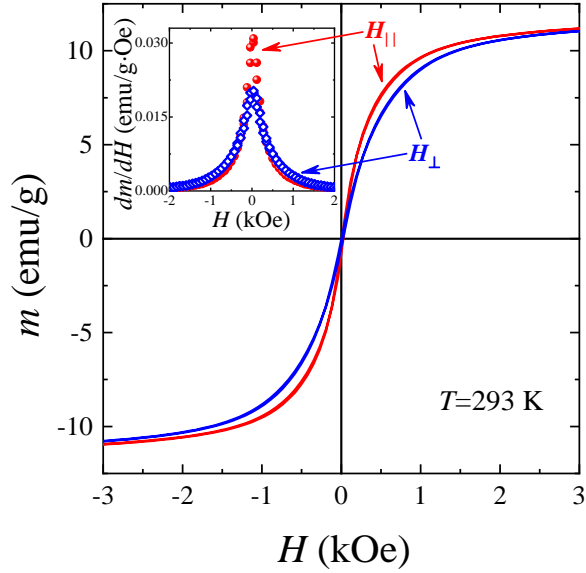

**Fig. 8**. Field dependences of the magnetization for the MAE beam along its axis $m_\parallel(H_\parallel)$ and perpendicular to its axis $m_\perp(H_\perp)$. The inset shows field dependences for the derivatives of $dm_\perp/dH_\perp$ and $dm_\parallel/dH_\parallel$.

The part of the beam with a volume of $\Delta V$ bent under the action of a magnetic field **H** is shown in Fig. 9. This small section of a beam is rotated in the field at an angle $\varphi$, and its magnetization vector **m** makes an angle $\theta$ with the direction **H**. The noncollinearity of vectors **m** and **H** is associated with the magnetic anisotropy of the beam. This leads to the act of the moment of forces $\Delta M = m\Delta VH \sin\theta$ on the $\Delta V$ of the MAE (see Fig. 9). The direction of the vector of the moment of forces **M** is determined by the vector product of **m** by **H**. Taking into account the anisotropy of magnetization, the modulus of magnetization is equal to $m = H\sqrt{\chi_\perp^2 \cos^2\varphi + \chi_\parallel^2 \sin^2\varphi}$. Thus, the magnitude of the moment depends on the field as $\Delta M = \Delta V H^2 \sin\theta \sqrt{\chi_\perp^2 \cos^2\varphi + \chi_\parallel^2 \sin^2\varphi}$. In low field, the field dependence for the moment of force is proportional to the square of the magnetic field. In stronger field, it becomes more complicated, since it is necessary to take into account the dependence of the angles θ and φ on *H*. This behavior of magnetically soft superparamagnetic particles differs from the behavior of magnetically hard particles, where force proportional to the first degree of the external magnetic field [59].

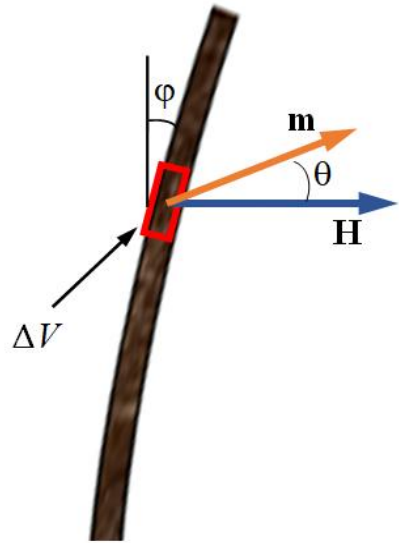

**Fig. 9**. Beam bending in a magnetic field due to the influence of magnetic anisotropy (the vectors of magnetization **m** and the field **H** are not collinear with each other).

Thus, there are two mechanisms of MAE bending in a magnetic field: i) the effect of inhomogeneity of the magnetic field and non-uniform particle distribution in MAE; ii) the effect of the magnetic anisotropy of the sample and, as a consequence, the appearance of the moment of forces Δ*M*. Both of these mechanisms lead to bending and their combined effect should be taken into account to explain the experiment.

It should be noted that in the case of a strongly pronounced magneto-rheological effect, when magnetic nanoparticles have the ability to displace in the matrix, the force Δ*F* leads to the MAE bending starting from the smallest values of the field *H*. If the initial distribution of particles is equally probable, the particles in the MAE do not shift and the force Δ*F* does not arise in a uniform magnetic field. In this case, only the moment of force Δ*M* generated during magnetization [72] will lead to the MAE bending. However, in a uniform magnetic field, the bending of a homogeneous MAE sample caused by a moment $\Delta M$ must develop critically. The sample will begin to bend when the magnetic field becomes greater than the threshold (critical) value [57]. If the particles are mobile and their non-uniform redistribution in the MAE volume occurs under the action of a weakly heterogeneous magnetic field, both force $\Delta F$ and $\Delta M$ actions act in the entire interval of fields.

The bending induced by the force $\Delta \mathbf{F}$ is proportional to the square of the field. Theefore, the bending force $\Delta \mathbf{F}$ does not change its sign when the direction of the magnetic field is reversed. This result is in good agreement with the experimental data (see Fig. 5), where the bending sign is not changed upon changing the direction of the magnetic field to the opposite. It is consistent with the data in Fig. 6, where $\Delta L$ is a function of $H^2$. The MAE sample will bend in the opposite direction (upon changing sing of $H$) if it has a remanent magnetization component directed along the axis of the beam. When the direction of $\mathbf{H}$ is reversed (acting on the remanent magnetization vector), the sign of the moment of forces $\Delta M$ and, as a consequence, the sign of bending $\Delta L$ can be changed for the magnetically hard particles.

## 4. Conclusions

The MAE with nanoparticles of nonstoichiometric $La_{0.8}Ag_{0.2}Mn_{1.2}O_3$ manganite with a Curie temperature $T_C = 308$ K slightly higher than room temperature was obtained. It was found that the $La_{0.8}Ag_{0.2}Mn_{1.2}O_3$ nanoparticles in MAE are magnetized superparamagnetically from room temperature to the Curie point. The coercivity in the MAE is about 10 Oe at room temperature. In the $La_{0.8}Ag_{0.2}Mn_{1.2}O_3$ nanopowder, the coercivity is higher due to the closer arrangement of particles from each other and greater opportunity to interact by magnetic dipole-dipole forces.

It was found that the beam bending does not change direction upon magnetization reversal. After magnetization, the beam has a residual bending that is associated with the non-equilibrium process of restructuring the positions of nanoparticles in MAE induced by the magnetic field.

The magnitude of the MAE bending in a magnetic field is determined by the superparamagnetic character of magnetization of the MAE nanoparticles and is described by a polynomial with 2-nd and 4-th powers in the field. The temperature dependence of the bending induced by magnetic field is qualitatively similar to the temperature dependence for the square of the magnetization of the MAE. In thermodynamic theory, such a deformation is a consequence of the competition between the elastic and magnetic torques of the magnetized MAE. A more complex nonlinear dependence of the bending $\Delta L(H)$ is associated with a change in the internal energy of the MAE due to restructuring.

It turned out that bending of MAE with nanoparticles can be observed starting from small magnetic field values. Nanoparticles are more mobile in the elastomer matrix. Therefore, MAE can be restructured even in a small and weakly inhomogeneous magnetic field. An increase in the concentration of magnetized nanoparticles at one of the sample surfaces due to restructuring creates an additional mechanical stress that compresses this surface, which promotes bending.

It has been shown that the magnitude of the MAE bending induced by the magnetic field has a temperature feature, which is anomalously (critically) changed near the Curie point $T_C$ upon transition to the paramagnetic state of the particles. While heating MAE above the Curie point, the magnetoelastic properties of nanoparticles disappear. While cooling below the Curie point, they are restored. Thus, the discovery of another new self-controlling process in MAEs may also be interesting from the applied point of view.

The revealed temperature feature of the magnetoelasticity of MAE with nanoparticles is of interest from the point of view of applied application. The effect of "on-off" of the magnetoelastic coupling at the Curie point can be used to control and manipulate the MAE response to external contactless action created by a magnetic field.

**Declaration of Competing Interest**

The authors declared that there is no conflict of interest.

**Data Availability**

Data available on request from the authors.

**CRediT authorship contribution statement**

**Yu.I. Dzhezherya**: Data curation; Investigation; Project administration; Writing - editing. **Wei Xu**: Investigation; Methodology. **S.V. Cherepov**: Data curation; Investigation; Formal analysis; Methodology. **Yu.B. Skirta**: Data curation; Formal analysis; Validation. **V.M. Kalita**: Conceptualization; Methodology; Project administration; Supervision; Roles/Writing - original draft. **A.V. Bodnaruk**: Data curation; Investigation; Formal analysis; Methodology. **N.A. Liedienov**: Data curation; Investigation; Project administration; Writing - editing. **A.V. Pashchenko**: Formal analysis; Data curation; Validation. **I.V. Fesych**: Data curation; Investigation; Methodology. **G.G. Levchenko**: Formal analysis; Project administration; Resources; Supervision; Writing - editing.

**Acknowledgements**

This work was partially supported by The Thousand Talents Program for Foreign Experts program of China (project WQ20162200339) and Grant of NAS of Ukraine for research laboratories / groups of young scientists of NAS of Ukraine in 2020-2021.